\documentclass[usenatbib]{mn2e}
\usepackage{graphicx}
\usepackage{lscape}

\newcommand{\gtsimeq}{\raisebox{-0.6ex}{$\,\stackrel
        {\raisebox{-.2ex}{$\textstyle >$}}{\sim}\,$}}

\author[Ben Burningham et al.]{Ben Burningham$^1$, Tim Naylor$^1$,
 S. P. Littlefair$^{1,2}$, R. D. Jeffries$^3$
 \\
$^1$ School of Physics, University of Exeter, Stocker Road, Exeter EX4 4QL\\
$^2$ Department of Physics and Astronomy, University of Sheffield, Sheffield
 S3 7RH \\
$^3$ Department of Physics, Keele University, Keele, Staffordshire ST5 5BG}

\title[Age Spreads]{Can variability account for
  apparent age spreads in OB association colour-magnitude diagrams?}

\begin{document}
%
%
%
%


\def\aj{\rm{AJ}}                   
\def\araa{\rm{ARA\&A}}             
\def\apj{\rm{ApJ}}                 
\def\apjl{\rm{ApJ}}                
\def\apjs{\rm{ApJS}}               
\def\ao{\rm{Appl.~Opt.}}           
\def\apss{\rm{Ap\&SS}}             
\def\aap{\rm{A\&A}}                
\def\aapr{\rm{A\&A~Rev.}}          
\def\aaps{\rm{A\&AS}}              
\def\azh{\rm{AZh}}                 
\def\baas{\rm{BAAS}}               
\def\jrasc{\rm{JRASC}}             
\def\memras{\rm{MmRAS}}            
\def\mnras{\rm{MNRAS}}             
\def\pra{\rm{Phys.~Rev.~A}}        
\def\prb{\rm{Phys.~Rev.~B}}        
\def\prc{\rm{Phys.~Rev.~C}}        
\def\prd{\rm{Phys.~Rev.~D}}        
\def\pre{\rm{Phys.~Rev.~E}}        
\def\prl{\rm{Phys.~Rev.~Lett.}}    
\def\pasp{\rm{PASP}}               
\def\pasj{\rm{PASJ}}               
\def\qjras{\rm{QJRAS}}             
\def\skytel{\rm{S\&T}}             
\def\solphys{\rm{Sol.~Phys.}}      
\def\sovast{\rm{Soviet~Ast.}}      
\def\ssr{\rm{Space~Sci.~Rev.}}     
\def\zap{\rm{ZAp}}                 
\def\nat{\rm{Nature}}              
\def\iaucirc{\rm{IAU~Circ.}}       
\def\aplett{\rm{Astrophys.~Lett.}} 
\def\apspr{\rm{Astrophys.~Space~Phys.~Res.}}
\def\bain{\rm{Bull.~Astron.~Inst.~Netherlands}} 
\def\fcp{\rm{Fund.~Cosmic~Phys.}}  
\def\gca{\rm{Geochim.~Cosmochim.~Acta}}   
\def\grl{\rm{Geophys.~Res.~Lett.}} 
\def\jcp{\rm{J.~Chem.~Phys.}}      
\def\jgr{\rm{J.~Geophys.~Res.}}    
\def\jqsrt{\rm{J.~Quant.~Spec.~Radiat.~Transf.}}
\def\memsai{\rm{Mem.~Soc.~Astron.~Italiana}}
\def\nphysa{\rm{Nucl.~Phys.~A}}   
\def\physrep{\rm{Phys.~Rep.}}   
\def\physscr{\rm{Phys.~Scr}}   
\def\planss{\rm{Planet.~Space~Sci.}}   
\def\procspie{\rm{Proc.~SPIE}}   

\let\astap=\aap
\let\apjlett=\apjl
\let\apjsupp=\apjs
\let\applopt=\ao

\linespread{1.3}
\maketitle

\begin{abstract}
We have investigated the role of photometric variability in causing
the apparent age spreads observed in the colour-magnitude diagrams of OB
associations.  
We have found that the combination of binarity, photometric
uncertainty and variability on
timescales of a few years is not sufficient to explain the
observed spread in either of the OB associations we have studied.  
Such effects can account for about half the observed spread in the
$\sigma$ Orionis subgroup and about 1/20 of the observed spread in Cep OB3b.
This rules out variability caused by stellar rotation and rotation of
structures within inner accretion discs as the source of the majority
of the  the apparent age spreads.    
We also find that the variability tends to move objects
parallel to isochrones in $V/V-i'$ CMDs, and thus has little influence
on apparent age spreads.
We conclude that the
remaining unexplained spread either reflects a true spread in the ages
of the PMS objects or arises as a result of longer term variability
associated with changes in accretion flow.

\end{abstract}


\section{Introduction}

A significant unresolved question in the study of star
formation is how long it takes.
  Whether we are considering the
entire process from the state of neutral interstellar hydrogen to the
Zero Age Main-Sequence (ZAMS), or just the portion of the process from
the fragmentation of a giant molecular cloud (GMC) onwards, there is no
consensus as to how long star formation takes.  
There are two
competing paradigms of star formation currently proposed in the
literature, which each give rise to very different timescales for star
formation.

\subsection{Slow star formation}

\citet{shu77} put forward the model of star formation which we
will refer to as slow star formation (SSF), which was reviewed by
\citet{sal87}.  
In this model, molecular cloud complexes are in dynamical equilibrium,
with lifetimes of several tens of megayears and are supported against
free fall collapse by magnetic fields.  
However, in clumps where the density is $n \gtsimeq 10^5$ cm$^{-3}$, the
ionisation fraction can be sufficiently low that the neutral molecular
material may diffuse through the magnetic field, removing flux from
the core (ambipolar diffusion),
leading to higher degrees of central condensation. 
Eventually the field can no longer support the, now prestellar, core
against gravitational collapse to form a hydrostatic protostar. 

Typically,  predicted ambipolar diffusion timescales lie in the
range 5-10 Myrs.  
This is consistent with the observed lifetimes of
pre-stellar cores with $10^5 < n < 10^6$ found by \citet{wtsha94} to
be $\sim 10^6$ yrs, through comparison of the number of starless cores
versus protostellar cores. 
However, \citet{jma99} performed a more comprehensive study of pre-stellar
cores using ammonia emission.  
They found the ratio of starless to stellar cores to be too small to be
consistent with that expected if the ambipolar diffusion timescale
governed core lifetimes. This suggests that the timescale of the
starless core phase is much less than the ambipolar diffusion timescale.

\subsection{Rapid star formation}

More recently, an alternative paradigm for pre-stellar core formation
and collapse has been discussed in the literature
\citep[e.g.][]{bphvs99,hbpb2001,hartmann2001,e2000}.  In this model the
star formation rate is regulated by supersonic turbulence, not
magnetic fields.  
Since supersonic turbulence in self-gravitating
clouds is expected to decay rapidly, the cloud support mechanism in this
picture leads to a short cloud lifetime compared to that in SSF
\citep{pal2001}.  So, rather than being long-lived structures, GMCs are
treated as transient objects that form through interactions between
supersonic flows in the interstellar medium.  
The whole process from formation of the GMCs through to the
arrival of the pre-main-sequence (PMS) stars on the birthline is expected to
take around 3 Myrs, with collapse of cores to form protostars occurring
almost immediately. As such we will refer to this model as rapid star
formation (RSF).  RSF has a number of distinct advantages over SSF,
both observationally and theoretically which are reviewed by \citet{mlk2004}.  

\subsection{Age spreads}

The presence of apparent age spreads has been observed in a number of
young clusters \citep[e.g.][]{hm82,sbl98} and
associations \citep[e.g.][]{pnjd2003,dm2001} and their sizes have been
presented as evidence in favour of both SSF and RSF.
\citet{ps2000} investigated apparent age spreads in a number of young
star forming regions and found evidence of accelerating star
formation, which they use to argue in favour of SSF.  
A further study by the same authors into the age spread in the
Taurus-Auriga region also found evidence of accelerating star
formation \citep{ps2002}.  
The evidence of accelerating star formation in the regions studied
by these authors came in the form of distributions of stellar ages
that were, generally, strongly peaked at 1 - 2 Myrs, with few PMS
stars older than this.   
As \citet{hartmann2003} pointed out, this surely implies there is something
special about the last 1-2 ~Myrs, if such widely separated regions
have formed the majority of their stars at the same time, whilst their
overall lifetimes are $\sim$ 10 Myrs. 
Such an uncomfortable, special, state of affairs is not required if
the one accepts the RSF paradigm. This is because the strong peaking of age
distributions at 1-2 Myrs follows naturally if this is the timescale
for cloud and star formation \citep{hartmann2003}.

\citet{e2000} found that the ages spreads seen in a number of OB
associations and clusters are comparable to the inferred crossing time
for the parent cloud.  
The conclusion that is drawn from this is that
the age spread is indicative of the timescale for star formation, and
that this is comparable to the crossing time, as expected for RSF.  
However, his data reveal that the age
spreads also scale with the mean ages for the groups he uses. 
This scaling of apparent age spreads with mean age for
clusters and associations suggests that the age spreads
originate from a photometric scatter of given magnitude. 
The size of this spread is much larger than any photometric uncertainties.  
Since PMS objects move more slowly through colour-magnitude
space at older ages, a given photometric scatter will naturally imply
larger age spread as the mean age rises. 
The correlation of spread with inferred crossing time may also arise as a
result of the fact that older open clusters and associations are
larger and thus a longer crossing time is inferred for the parent
cloud.

\citet{hartmann2001, hartmann2003} disputed the results and conclusions of
\citet{ps2000,ps2002}, arguing that the actual age spreads are much
smaller than those observed, but accepted that the presence of such
large age spreads ($10^7$ yrs) would be a problem for the RSF paradigm.
Additionally, it is not
clear that the interpretation of the age spreads being indicative of
timescale is correct. 
\citet{tcm2004} pointed out that interpretation of age spreads in this
manner already assumes a core formation timescale that is essentially
instantaneous with respect to the lifetime of the molecular cloud,
$\tau_{mc}$. 
In reality, an age spread could only tell us the difference between
$\tau_{mc}$ and the timescale for forming a body that will survive the
destruction of the molecular cloud, the core formation timescale,
$\tau_{cf}$.  
So: $\tau_{spread} = \tau_{mc} - \tau_{cf}$.  As such, interpretation of
any spread in ages is dependent on knowledge of either $\tau_{cf}$ or
$\tau_{mc}$. 

Incidentally, this point is also relevant to the argument put forward by
\citet{hbpb2001} that, since very few young clusters or associations
older than 3 Myrs are associated with their parent cloud, star
formation must occur on a timescale of about 3 Myrs.  
This also implicitly assumes that $\tau_{cf}$ is short.  
The same observational evidence
could indicate that star formation is slow, but that the cloud is
disrupted quickly after the first PMS stars arrive at the birthline.
However, as pointed out by \citet{hbpb2001}, the vast majority of
local molecular cloud complexes show evidence of star formation.  
This must imply that $\tau_{cf}$ should be short, or the question must surely
be asked as to the whereabouts of the clouds which are still mostly in
the pre-stellar phase of star formation.  
This is reflected in the results of \citet{jma99}, described earlier,
that the ratios of pre-stellar to stellar cores found in molecular clouds are
far below the 3 to 30:1 range required by SSF.

It is important to recognise, however, that the reality of
 apparent age spreads has not been well established.  
The principal evidence for spreads of ages actually comes from an
observed spread of PMS stars in colour-magnitude (C-M) space.  A
single age would be expected to give rise to a much more narrow
distribution of stars about an isochrone.
There are a number of plausible
alternatives that could explain such C-M spreads for a
population that in reality arrived at the birthline simultaneously.
For example accretion-driven age spread \citep{tlb99} could give
rise to an apparent spread of ages for an ensemble of objects with
differing accretion histories.  
\citet{hartmann2001} explored a number of other factors that might be
expected to influence the degree of observed age spread.  
These included variable
extinction, photometric variability, differences in accretion
luminosities and the presence of unresolved binaries.
 Establishing the influence of
variability and binarity on the width of the PMS in OB association-like
environments is the aim of this work.  We have obtained 2 epoch, 2
colour photometry for 2 regions with differing photometric spreads to
investigate this effect. 
The 2 regions studied here are within well known OB associations: Cep
OB3b and the $\sigma$ Orionis young group (part of the Orion OB1b association).
 We have simulated the degree of spread
introduced by variability on timescales of less than 1 year in Cep
OB3b and less than 4 years in $\sigma$ Ori.  By comparing
2 colour catalogues of PMS objects within these associations we have
estimated their variability, and used this to simulate the PMS in C-M
space, assuming a single isochronal age for the objects.

The rest of the paper is laid out as follows.  In section
~\ref{sec:spreads_obs} we describe our observations and data reduction.  In
section ~\ref{sec:sims} we describe how we have simulated the spreads
in each of our associations in turn, and give the basic results.
These results are discussed in section ~\ref{sec:spreads_disc}, and our
conclusions are summarised in section ~\ref{sec:spreads_concs}.

\section{Observations and Data Reduction}
\label{sec:spreads_obs}

All observations were carried out using the Wide Field Camera (WFC) mounted
on the Isaac Newton Telescope (INT) at the Roque de los Muchachos
Observatory, La Palma.  Our first epoch data set for
$\sigma$ Ori is the same as that presented by \citet{kenyon2005}.  It
was made up of observations taken on the nights of 27-30
September 1999 of 5 fields of view (FoV) in the Harris $R$ and Sloan $i'$
filters.  We carried out observations of 4 FoVs coincident with Kenyon
et al's survey on the night of 7 September 2003, using the same
filters.  The new observations are detailed in Table ~\ref{table:obs}.  We
have obtained new data for both epochs for our Cep OB3b survey.  We
have observed a single WFC FoV in this region, chosen to cover the
area with the highest density of PMS objects identified by
\citet{pnjd2003}.  Observations were taken on the nights of 12 September
2003 and 28 September 2004 using Harris $V$ and Sloan $i'$ filters.  Again
the observations are detailed in Table ~\ref{table:obs}.

\begin{table*}
\caption{Summary of Observations}
\label{table:obs}
\begin{tabular}{c c c c c c c c c c c}
\hline
Field Name & 
\multicolumn{2}{ c } {$\sigma$ Ori 1} &
\multicolumn{2}{ c } {$\sigma$ Ori 2} &  
\multicolumn{2}{ c } {$\sigma$ Ori 3} &
\multicolumn{2}{ c } {$\sigma$ Ori 4} & 
\multicolumn{2}{ c } {Cep OB3b} \\
\hline
RA(J2000) & 
\multicolumn{2}{ c }{05 40 14.2} &
\multicolumn{2}{ c }{05 40 13.1} &
\multicolumn{2}{ c }{05 38 07.7} & 
\multicolumn{2}{ c }{05 38 07.4} & 
\multicolumn{2}{ c }{22 55 43.3} \\
Dec(J2000) &
\multicolumn{2}{ c }{-02 20 18.1} &
\multicolumn{2}{ c }{-02 51 48.0} &
\multicolumn{2}{ c }{-02 20 18.0} &
\multicolumn{2}{ c }{-02 51 51.0} &
\multicolumn{2}{ c }{+62 40 13.7} \\
\hline
\hline
Filter & R & i' & R & i' & R & i' & R & i' & V & i' \\
\hline
2003-09-07 & 600s & 300s & 600s & 300s & 600s & 300s &
600s &  300s & - & - \\
2003-09-12 & - & - & - & - & - & - & - & - & 30s, 2s & 13s, 2s \\
2004-09-28 & - & - & - & - & - & - & - & - & 30s, 5s & 300s, 30s, 5s \\

\hline
\end{tabular}
\end{table*}

Data obtained in 2003 and 2004 were reduced in an identical manner.
Flatfields and data frames were linearised using the 2003 August
coefficients (see
http://www.ast.cam.ac.uk/$\sim$wfcsur/technical/foibles/index.php for
details), and then bias subtracted using a median bias frame specific
to each night.  We flat
fielded the data using frames constructed from twilight sky flatfields
taken during the same observing runs as the data being corrected.
The $i'$-band frames were successfully defringed using a fringe frame from 2001
September, obtained from the web pages of the Cambridge Astronomy
Survey Unit (CASU).

Optimal photometry was performed using the method laid out by
\citet{n98} and \citet{ntjpdt2002}, with the revisions described in
\citet{me2003} and \citet{littlefair2005}.  We have allowed the profile
correction to vary with position on each chip, and fitted it with a 3rd
order polynomial in the x-axis and a 5th order polynomial in the
y-axis. 
We do not apply the `ill-determined
sky' flag for the Cep OB3b data, and we allow the thresholds for its
application to vary from field to field for the $\sigma$ Ori data. 
For obvious reasons we did not apply the `variable' data quality flag
to reject objects that showed evidence of variability.
An astrometric solution was obtained through comparison with a 2MASS
catalogue for each FoV.  The RMS of the residuals to the 6-coefficient
fits were all less than 0.1 arcsec.

As described by \citet{ntjpdt2002}, overlap regions between pointings were
used to normalise the catalogues for each pointing onto a single
system. 
We use the same method to bring observations from 2 epochs for each of
our target regions to the same system.  
In the case of the $\sigma$ Ori data, the way in which flatfields were
normalised had changed between the two epochs, so the data from each
epoch were treated slightly differently.  When the 1999
data were reduced, the flatfield for each chip on the WFC was
normalised separately.  By the time the 2003 data were
reduced, the reduction software had been changed such that the
flatfields for all 4 chips in each pointing were normalised together.  
As such, prior to catalogue normalisation, the 1999 exposures were combined to
produce 1 catalogue for each chip, whilst the 2003 exposures were
combined to produce 1 catalogue for each pointing.  The overlap
regions between the interlocking pointings were then used to normalise
the catalogues onto a single system.

To verify that there were no
major sources of error introduced at the image processing stage we
checked that the distribution of positive and negative differences
between observations were uniform with position on the sky for the
$\sigma$ Ori data.  
We found
that the trends in the differences
between observations were correlated with the pixel coordinates on
each chip from the 1st epoch catalogue.  \citet{kenyon2005} found the RMS
of differences between overlap stars seen in their data to be 0.05 mags
and speculated that it was due to problems with the flatfield, which
is consistent with what has been seen here.  They state that the camera was
suffering from light leaks during their observing run, and it is
likely that this is the origin of the problem. 
We fitted the trends in the differences in x
and y for each chip with 1st and 2nd order polynomial functions, and
corrected the magnitudes of the objects in the 1st epoch catalogue
accordingly, prior to re-normalising the 2 epochs.
The RMS of the differences in the magnitudes in the
interlocking overlap regions after normalisation suggested that an
additional magnitude independent uncertainty of about 0.01 mags was
also present.  As in \citet{ntjpdt2002}, this uncertainty was included
in the uncertainty estimate for all objects.
Following this we
found that positive and negative differences for the $\sigma$ Ori data
were spatially uncorrelated.

Since no overlap regions were present in the Cep OB3b data sets, we
could not use the results of the normalisation to determine the size of any
additional uncertainty, so we applied the same value as found for the
$\sigma$ Ori data set. The distribution of positive and negative
differences were found to be spatially uncorrelated in the Cep OB3b data set.

We have not applied a measured photometric calibration to our final
catalogues as this would have involved transforming our Harris and
Sloan magnitudes and colours into the Cousins system.  
Such a transformation would risk
introducing spurious correlations between variability and colour,
whilst not improving our experiment in any way.  As such we have
simply applied mean zero points to our magnitudes and colour
coefficients to our colours to construct our CMDs.

\section{Simulation of photometric spreads}
\label{sec:sims}

\subsection{The $\sigma$ Ori subgroup}
\label{sec:sim_sori}

\begin{figure}
\includegraphics[height=375pt,width=275pt, angle=90]{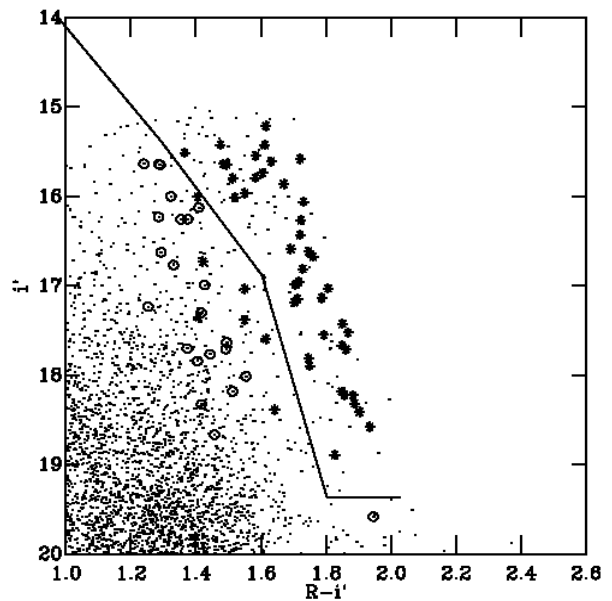}
\caption{The CMD for the 2nd epoch $Ri'$ catalogue. Spectroscopic members
 of \citet{kenyon2005} and \citet{me2005} are shown as asterisks.
 Objects identified as non-members in the same studies are shown as
 open circles.  The solid line indicates the location of the cut for
 our PMS selection.}
\label{fig:sori_memsel}
\end{figure}

The $\sigma$ Ori young group has been the subject of a number of
spectroscopic studies aimed at identifying bona fide members. In
particular two campaigns \citep{kenyon2005, me2005} were successful in
identifying a large number of members present in the catalogue obtained
from the 1st epoch observations.  Importantly, these spectroscopic
surveys also demonstrated that photometric selection of the PMS
objects in this region is not subject to serious contamination, and
does not miss significant numbers of bona fide members.  As such we
have used the membership lists from these studies to define a PMS
region in the CMD, with minimal contamination, from which we have
drawn our sample (see Figure~\ref{fig:sori_memsel}).  
Only objects with little doubt as to their membership have been used to
guide the photometric selection.  
In the case of the \citet{me2005}
catalogue this meant objects with greater than 90\% membership
velocity probability. 
In the case of the \citet{kenyon2005} catalogue this meant objects
which displayed strong Li absorption, evidence of low surface gravity
and the appropriate radial velocity.

To test if the observed C-M spread is affected by variability
we have assumed a single age for the young group and then simulated the PMS
using an estimate of each object's variability, derived from our 2
epoch observations.  This method implicitly includes the effect of
photometric uncertainty.

\begin{figure}
\includegraphics[height=250pt,width=190pt, angle=90]{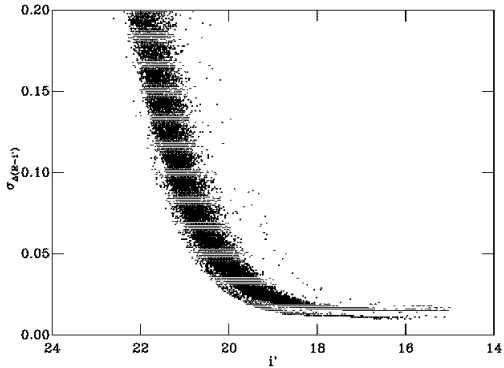}
\caption{A plot of the uncertainty in $\Delta (R-i')$ against $i'$ for the
 FoV centered on the $\sigma$ Ori subgroup.}
\label{fig:sori_errors}
\end{figure}

To verify that our PMS selection is indeed more variable than the
background we have compared the RMSs of the differences between the two sets
of observations for the background objects and for the PMS selection.
To make such a comparison meaningful we have restricted our sample to those
$i'$ magnitudes where the uncertainties are small.  
In Figure ~\ref{fig:sori_errors} we plot the uncertainty in the $R-i'$
differences (thus incorporating uncertainties from both epochs)
against $i'$ (2nd epoch).  
As can be seen, the uncertainties rise sharply fainter than $i'= 18$,
as the uncertainty in the photon counting statistics of the sky begins
to dominate.
Brighter than $i' = 16$, some objects also display higher
uncertainties.  
This is because for these points the 1st epoch data were drawn
mainly from a short exposure. 
 In the range 15 $< i' <$ 18 the
uncertainties are dominated by the small magnitude independent uncertainty
measured during the normalisation of the catalogues
\citep[see][and Section \ref{sec:spreads_obs}]{ntjpdt2002}. 
Based on this plot we restrict our comparison of the RMSs
to the range 15 $< i' <$ 18.  
The RMS of the differences between the two sets of observations
indicate that the objects in the photometric PMS selection are
significantly more variable than those in the background.  
The RMS for the differences in
$i'$ are 0.05 for the background and 0.09 for the PMS region.  In $R-i'$
the RMSs are smaller: 0.02 for the background; 0.04 for the PMS
region.

We placed the objects on a single empirical isochrone by fitting
a 3rd order polynomial to the photometrically selected PMS in the CMD.  
Each point used for fitting the polynomial is weighted according to
its uncertainty in $R-i'$, with the points carrying the greatest
uncertainty having the smallest influence on the fit.   
A point was
then placed on this empirical isochrone at the appropriate magnitude
for each PMS object.
One can view this as moving each point in colour until it falls on
the isochrone.

Having placed each point on the isochrone, we simulated the effect
of binaries by splitting this empirical isochrone into a single star and an
equal mass binary sequence.  
We did not assume a binary fraction, but
rather simulated spreads with a range of fractions.  For each binary
fraction we randomly selected that proportion of objects to be offset in
magnitude by an amount equal to $0.75(1-f_{bin})$, and the rest by
$0.75 - 0.75(1-f_{bin})$.  This gave rise to a difference of
0.75 mag between the two sequences, as would be expected from equal
mass binaries.  We do not attempt to simulate $q < 1$ binaries as
these would fall within the envelope bounded by our two sequences, and
thus would not have a significant impact the results of this investigation.

To simulate the effect of variability we next moved each point by
 an amount in both magnitude and colour equal to 
 $(\Delta / \sqrt2)$, where $\Delta$ is the observed difference between the two
observations for that point.  
Characterising the variability in this manner for each object in turn
has distinct advantages over parameterising the scatter of the
whole sample, as any correlations between $\Delta i'$ and $\Delta (R-i')$
have been included without any assumptions as to the source of the
variability.   Furthermore, any correlation of variability with
 magnitude has also been included, as has the influence of photometric
 uncertainty.  

\begin{figure}
\includegraphics[height=250pt,width=190pt, angle=90]{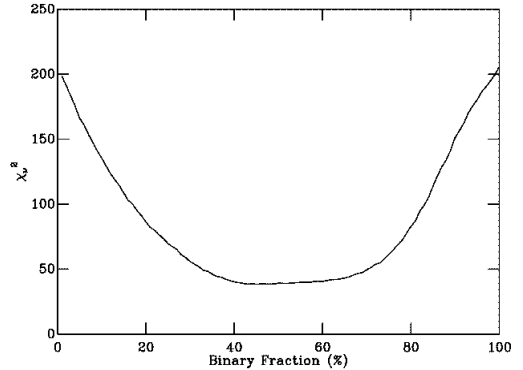}
\caption{The mean $\chi_{\nu}^2$ for the simulated PMS
  compared to the observed PMS plotted against binary fraction, shown as a
  percentage for the PMS objects near $\sigma$ Ori.}
\label{fig:sori_bf}
\end{figure}

\begin{figure*}
\includegraphics[height=200pt,width=500pt, angle=0]{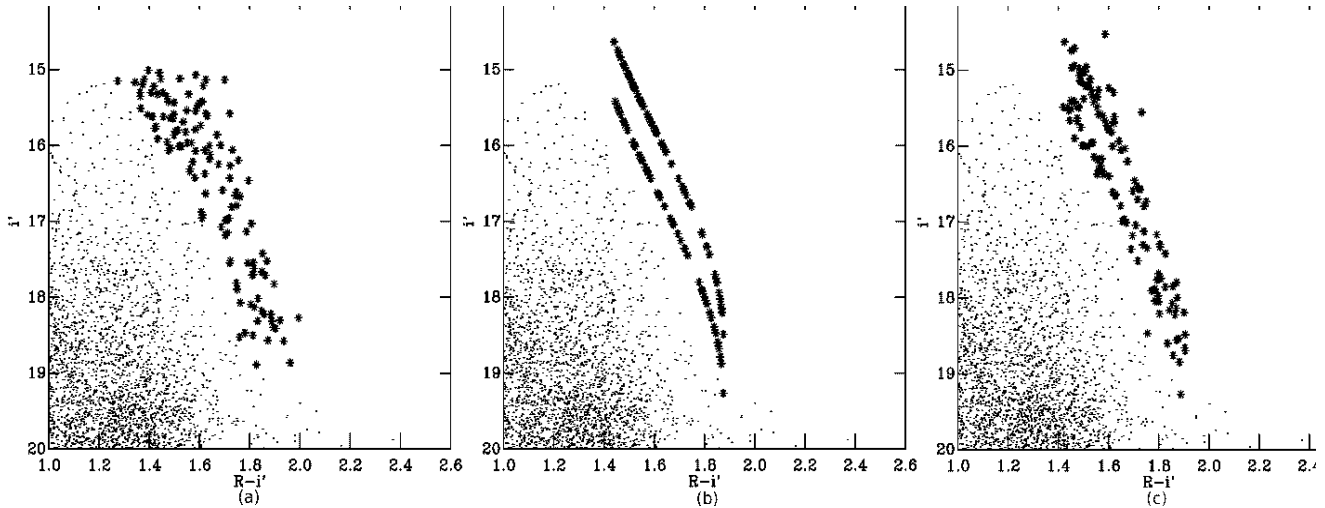}
\caption{CMDs for the $\sigma$ Ori young group showing: (a) the PMS
  selected objects as asterisks; (b) the fitted single star and binary
  sequences for a binary fraction of 50\%; (c) the simulated PMS for
  the same binary fraction.}
\label{fig:sori_results}
\end{figure*}

A further source of scatter that might influence the observed spread
is differential reddening.
We have neglected the effects of differential reddening in this
simulation since the extinction towards $\sigma$ Ori is low,
with a colour excess of just $E(B-V) = 0.05$ \citep{lee68}.
To select which binary fraction gave the best match to
the data, and thus which one would be used for subsequent analysis, we
constructed a $\chi_{\nu}^2$ estimate for each simulation by comparing
a histogram of the residuals in $R-i'$ from the polynomial fit to the
data, with a histogram of the residuals from a polynomial fit to the
simulation.
The histograms were all constructed in an identical manner: we placed the
residuals in bins of width 0.05~mags starting at -0.5 and ending
at +0.5.
Because we used a random number generator to select the objects that
 make up the binary sequence, we ran each simulation 1000 times
 to obtain a mean value for $\chi_{\nu}^2$, thus reducing the noise for the
 determination of the most likely binary fraction.  Figure
 ~\ref{fig:sori_bf} shows a plot of mean $\chi_{\nu}^2$ against binary
 fraction.  Clearly there is no sharp minimum in the value of
 $\chi_{\nu}^2$, but a broad minimum is centered on 50\%.  We present
 one realisation of the simulation for a binary fraction of 50\% in
 Figure ~\ref{fig:sori_results}.

\begin{figure}
\includegraphics[height=250pt,width=190pt, angle=90]{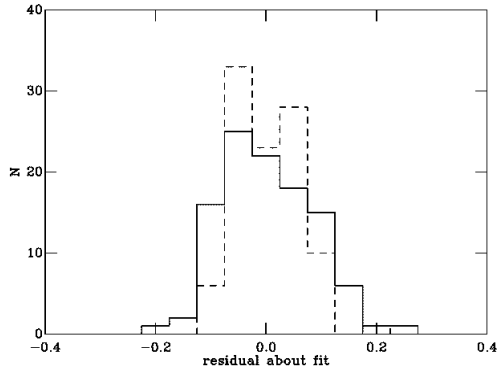}
\caption{Histograms of the residuals in $R-i'$ about polynomial fits to the
  observed PMS (solid line) and the simulated PMS (dotted line) for
  the $\sigma$ Ori PMS sample (binary fraction 50\%).}
\label{fig:sori_reshists}
\end{figure}

Comparison of panels (a) and (c) in Figure~\ref{fig:sori_results}
indicates that the combination of variability and binarity are not
able to account for the spread in the PMS.  
In Figure~\ref{fig:sori_reshists} we show
histograms of the residuals about 3$^{rd}$ order polynomial fits to the
observed PMS and the simulated PMS (solid and dotted line) for the $15
< i' < 18$ region of the CMD. Again, it is clear that the combination
of binarity and variability on timescales of $\sim4$ years, are not sufficient
to explain the spread in C-M space for this PMS.
With such a poor match to the observed spread it is clear why Figure
~\ref{fig:sori_bf} displays no sharp minimum in $\chi_{\nu}^2$, but
rather a broad minimum centered on 50\%.  
A binary fraction of 50\%
maximises the size of the spread, for a given estimate of variability. 
The assumption of a $q = 1$ binary population also increases the value of
$\chi_{\nu}^2$ as it introduces a double peak to the spread which is
not present in the data.

The FWHM of the spread in residuals in
$(R-i')$ about the polynomial fit to the observed PMS, shown in Figure
~\ref{fig:sori_reshists} is approximately 0.3 mags.
This is consistent with the width of the PMS
observed by \citet{sww2004} in $V-I$ for more massive members of the
same group.  

\subsection{Cep OB3b}
\label{sec:sim_cep}

\begin{figure}
\includegraphics[height=375pt,width=275pt, angle=90]{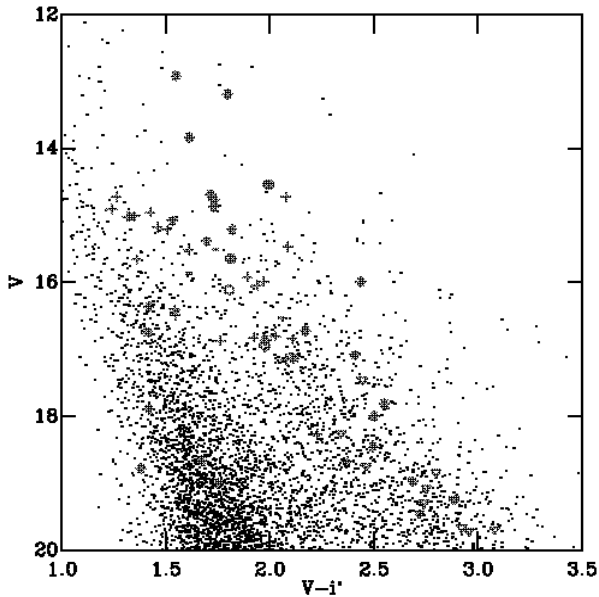}
\caption{The CMD for the 1st epoch Cep OB3b data.  Members identified
  by \citet{osp2002} are shown as open triangles, those from
  \citet{nf99} are shown as asterisks, CTTSs from \citet{pnjd2003} are
  shown as open squares, while the WTTSs are shown as open circles.
  Confirmed non-members from \citet{pnjd2003} are shown as crosses.
  Those WTTS that were identified in both \citet{nf99} and
  \citet{pnjd2003} are shown as filled circles.}
\label{fig:cep_memsel}
\end{figure}

In this case we do not have the benefit of such a large 
spectroscopic sample as in the $\sigma$ Ori group. As a result,
we have no estimate of the likely contamination from field stars, or
the true extent of the PMS region in C-M space.  There have, however,
been a number of studies that have identified likely low-mass members using
a variety of techniques. \citet{osp2002} used H$\alpha$ spectroscopy
to identify classical T-Tauri stars (CTTSs) in the vicinity of bright
rimmed clouds, and found 33 likely members of Cep OB3b, of which 16
are identified in our catalogue.  \citet{nf99} used ROSAT
observations to identify 56 X-ray sources toward Cep OB3b using
both HRI and PSPC observations.  We have cross correlated
their X-ray catalogue with our optical catalogue, matching the
brightest star within a 
radius of 14" for the PSPC positions and 7" for those from HRI.  We
find 21 PSPC objects correlate with objects in our catalogue, and 14
HRI objects.  
 We reject all objects
that lie in the Galactic background region on the left of the CMD
from further use in this study as they are likely non-members  \citep{me2005},
identified by chance correlations with our catalogue.
We accept as likely members those X-ray sources that correlate
with objects in the expected PMS region of the $Vi'$ CMD shown in
Figure ~\ref{fig:cep_memsel}.
\citet{pnjd2003} used radial velocities and Li {\sc i} absorption to identify 
both CTTSs and Weak T-Tauri stars (WTTSs) in Cep OB3b, and found 5
CTTS members and 5 WTTS members.  
Of these, all of the CTTSs and all-bar-one of the WTTS
are found in our catalogue.  \citet{pnjd2003} were also able to rule
out membership for a number of objects and these are also indicated on
the CMD.  
  Figure ~\ref{fig:cep_memsel} shows the
CMD for the first epoch data, with members from each survey overlaid,
along with non-members from \citet{pnjd2003}. 
Since we are unable to make a reliable PMS selection beyond those
members selected by previous authors, we simulate the
photometric spread for these 49 likely members.  
Since more than half our
sample has been selected on the basis of X-ray activity, it might be
argued that our sample is biased in favour of objects with the
greatest rotational variability: WTTSs.  
On the other hand, WTTS variability is smaller in magnitude than
non-periodic CTTS variability.  
As such, our simulation should still provide an indication of the
contribution from variability to the observed spreads.

\begin{figure}
\includegraphics[height=250pt,width=190pt, angle=90]{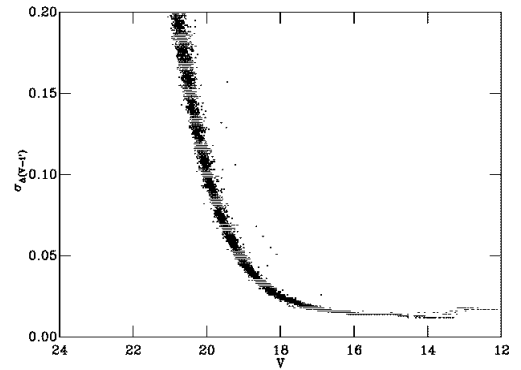}
\caption{A plot of the uncertainty in $\Delta (V-i')$ against V for the
 FoV in Cep OB3b.}
\label{fig:cep_errors}
\end{figure}

We have verified the variability of our member sample for Cep OB3b in the
same manner as for the $\sigma$ Ori sample.
  Figure ~\ref{fig:cep_errors} shows the
uncertainty in the $(V-i')$ differences as a function of V (1st epoch).  It
is clear that the uncertainties start to increase dramatically fainter than
about V = 18, and step up slightly at brighter than V = 13 for reasons
similar to those described in Section ~\ref{sec:sim_sori}.  
As such we have calculated
the RMSs for objects in the range of magnitudes 14 $< V <$ 17.5.  The RMS of
the differences between the two sets of observations indicates that the
likely members are significantly more variable than the rest of the
sample.  The RMS of the differences in $V$ are 0.07 for the total
sample, compared with 0.11 for the likely members.  In $(V-i')$ the
RMSs are smaller: 0.03 for the total sample; 0.07 for the likely members. 
A similar result is obtained if we attempt to make a photometric PMS
selection. 
 This indicates that a high proportion of objects in this region of
 the CMD may also be PMS members of Cep OB3b.

\begin{figure}
\includegraphics[height=250pt,width=190pt, angle=90]{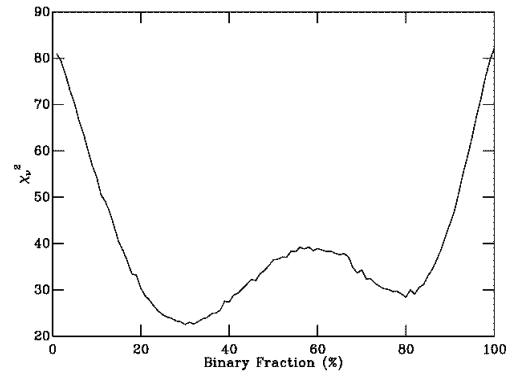}
\caption{The mean $\chi_{\nu}^2$ for the simulated member sequence
  compared to the observed member sequence plotted against binary
  fraction, shown as a percentage for likely members of Cep OB3b.}
\label{fig:cep_bf}
\end{figure}

The member sequence was simulated in the same manner as the
PMS sample in the previous section.  
  As before, we were able to neglect
the effect of differential reddening in our simulation, but for a
different reason.  
\citet{pnjd2003} found that the reddening vector in a $V/V-i'$ CMD lies
nearly parallel to the PMS for sight-lines toward Cep OB3b.  As such
the differential reddening is unlikely to add any spread to the
observed sequence.
  The only difference between the
two methods is that we have used a different bin size and range when
constructing the histograms used for determining $\chi_{\nu}^2$. 
Because the sample size is smaller we used a larger bin width (0.1),
whilst extending the range of the bins (-2.0 - +2.0) to include some
larger residuals.   As can be seen in Figure ~\ref{fig:cep_bf}, a
binary fraction of 25\% gives the closest match to the data.

\begin{figure*}
\includegraphics[height=200pt,width=500pt, angle=0]{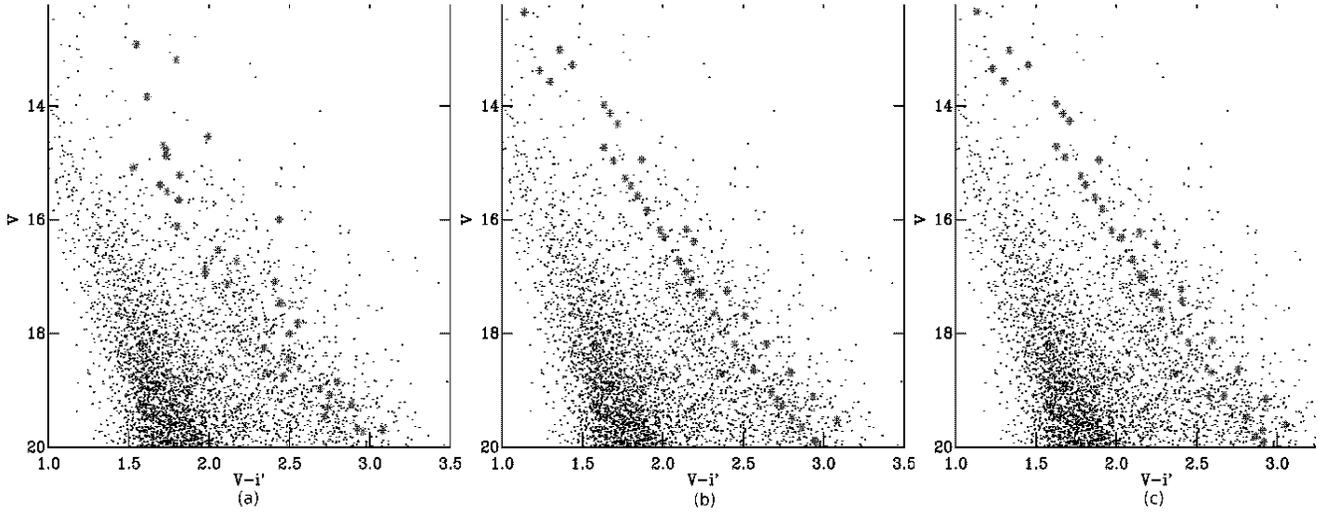}
\caption{A CMD showing the fitted single star and binary member sequences as
  lines of red asterisks for a binary fraction of 0.25 for the Cep OB3b.}
\label{fig:cep_results}
\end{figure*}

\begin{figure}
\includegraphics[height=250pt,width=190pt, angle=90]{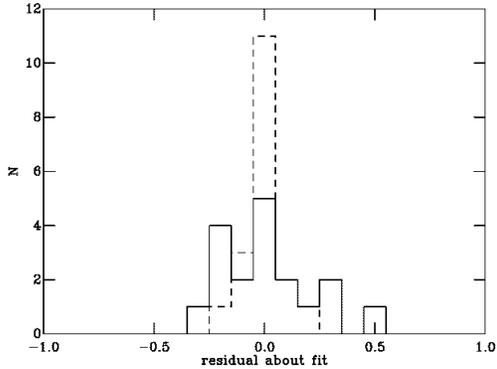}
\caption{Histograms of the residuals in $V-i'$ about polynomial fits to the
  observed member sequence (solid line) and the simulated member
  sequence with a binary fraction of 25\% (dotted line) for Cep OB3b.}
\label{fig:cep_reshists}
\end{figure}

The fitted member sequence 
for this binary fraction is shown overlaid on CMD (b) in Figure
  ~\ref{fig:cep_results}, whilst  the simulated member sequence is
shown on CMD (c) of the same figure.  In Figure
~\ref{fig:cep_reshists} we have plotted  a histogram of the
residuals about the fit to the data and the simulation for the $14 < V
< 17.5$ region of the member sequence for one realisation of the
simulation, using a binary fraction of 25\%.  Inspection of Figures
~\ref{fig:cep_results} and ~\ref{fig:cep_reshists} indicates that, as
was seen in the previous section for the $\sigma$ Ori young group,
the combination of binarity and variability on a timescale of $\sim$ 1 year
is not able to explain the spread of members in C-M space.

\section{Discussion}
\label{sec:spreads_disc}

We have shown that photometric variability on
timescales of 1-4 years is not able to account for the spread in
C-M space occupied by either of our sequences. 
If we assume that the influence of variability and binarity are added
to the underlying distribution in quadrature, we can estimate the
proportion of the observed spread that is accounted for here and that
which remains unaccounted for.
In the case of the
$\sigma$ Ori young group it appears that short term variability can
account for about half of the observed spread in $Ri'$ C-M space.  The
RMS of the residuals about the polynomial fit is 0.084 mags for the
observed PMS, and 0.057 mags for the simulated sequence, which leaves 0.062
mags unaccounted for.  Short term
variability can only account for about 1/20 of the spread in Cep OB3b,
where the RMSs of the residuals about the polynomial fits are 0.20 mags
for the observed member sequence, and 0.075 for the simulated
sequence, leaving 0.185 mags unaccounted for.

\subsection{The nature of the variability}

\begin{figure}
\includegraphics[height=225pt,width=200pt, angle=90]{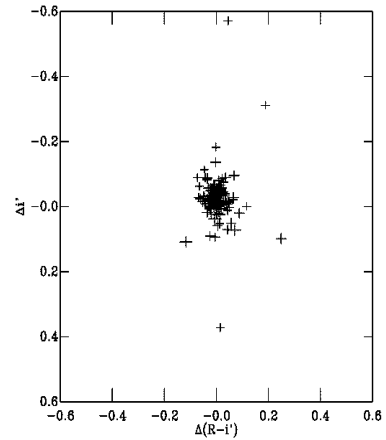}
\caption{A plot of $\Delta i'$ against $\Delta (R-i')$ for the $15 < i' <
  18$ region of the PMS in $\sigma$ Ori.  
The points are plotted with error bars.}
\label{fig:sori_diffs}
\end{figure}

\begin{figure}
\includegraphics[height=225pt,width=200pt, angle=90]{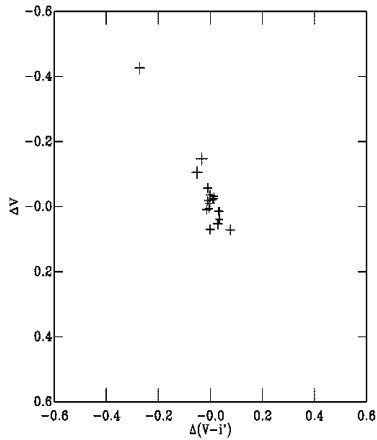}
\caption{A plot of $\Delta V$ against $\Delta (V-i')$ for the $14 < V <
  17.5$ region of the member sequence of Cep OB3b.  The points are
  plotted with error bars.}
\label{fig:cep_diffs}
\end{figure}

 Figures
~\ref{fig:sori_diffs} and ~\ref{fig:cep_diffs} show the 
measured differences in
colour and magnitude for the objects in our final sample for each of
our regions of interest.   The distribution of C-M shifts is clearly
different for the two regions, which is not surprising considering the
different colours used for each case.   
The diagonal distribution of the Cep OB3b differences demonstrates that
the variability seen here will only tend to move objects up and down
an isochrone, and will have little impact on the apparent age spread.  
These differences, which indicate that objects get bluer when brighter, are
consistent with variability arising from hot or cool
spots on the surfaces of PMS stars rotating in and out of view. 
The differences shown in the plot for the $\sigma$ Ori
subgroup display an uncorrelated distribution, which will tend to
displace objects across isochrones more.  As a result, the simulations
carried out for this region display a greater apparent age spread.
Since the $\sigma$ Ori observations are in $Ri'$, they
are less sensitive to colour changes associated with rotational
variability than $Vi'$ observations \citep{hhgw94}, and so the
different distributions of differences do not necessarily imply a
different origin.

Our observations rule out variability on timescales of 1-4 years as
the entire cause for the observed C-M spreads.
The mechanisms this therefore excludes are
rotation of hot or cool spots, and short timescale
non-periodic T-Tauri variability, such as that caused by
accretion noise or chromospheric flaring.
We can also rule out variability associated with rotation of
structures in the disc out to a radius of 1 A.U as a source for the
C-M spread, since the rotation of such structures should occur within
a year.  It is also unlikely that rotation of bright structures beyond 1
A.U would be responsible for the C-M spread as the temperature in a
thin disc drops rapidly with distance, such that at a radius of 1 A.U
the temperature is $\sim$ 100 K.  As such the outer disc is only
likely to contribute appreciable flux in the far-IR.

The RMS of residuals about the
polynomial fit to the simulated PMS is larger in Cep OB3 than the $\sigma$
Ori young group. 
Since accreting objects are known to be more variable that
non-accreting objects, this is consistent with the observed incidence
of accretors in the two regions.  
\citet{kenyon2005} found the fraction of low-mass accretors to be $10
\pm 5 \%$ in the $\sigma$ Ori young group, whilst the small number
of members identified by \citet{pnjd2003} in Cep OB3b suggest about
50\% of objects there are accretors.

Since longer timescale variability may be the origin of the remaining
unexplained spread we cannot confirm the reality of the apparent age
spreads.  However, what is clear is that the actual size of
either age spread is smaller than that observed, and possibly zero.  
Returning to the RMSs of the residuals about the polynomial fits, we
recall that the unaccounted for spread
has an RMS of residuals about the fit of 0.062
mags in the case of the $\sigma$ Ori subgroup, and 0.185 mags in the
case of Cep OB3b.

It is tempting read much into the observation that
the larger remaining spread is seen in the association whose natal
molecular would have had the larger crossing time.
However, it should be noted that in the case of the $\sigma$
Ori subgroup the apparent spread, and thus the underlying one also,
may be larger since we were conservative in our PMS selection.  
As can be seen in Figure ~\ref{fig:sori_memsel} we have actually
excluded some members in order to avoid risking significant
contamination by non-members in our photometric selection.
The excluded members have the same mean variability as the
members that were selected, so the simulated spread has not been
altered significantly by their exclusion.
  
Although we are confident that our selection is not subject to
significant contamination \citep[see][]{me2005,kenyon2005}, there
will, none-the-less, be some non-members included in the sample. 
These objects will not increase the size of the observed
spread since our confirmed members span the entire selected region of
the CMD (see Figure ~\ref{fig:sori_memsel}).  
However, it is likely that they will reduce the RMS of residuals about
the polynomial fit to the
simulated spread, as any non-members can be expected to be less
variable than the members.
If this is the case, then it may still be that the remaining unexplained
spread in the $\sigma$ Ori is very small.

It is still possible that the C-M spreads arise as a result of
some other kind of photometric variability.
Such variability could arise from variations in
accretion flow caused by stellar magnetic cycles
\citep[e.g.][]{armitage95}, with time-scales of a few years to
decades. Alternatively, much longer time-scale variability
such as that resulting from accretion-driven age spreads (ADAS) could
be to blame. 
Fundamentally, the influence of such long term accretion processes on the
presence of apparent age spreads would be best determined through a
study of accretion rates across a large number of objects in
associations that display a range of apparent spreads, rather than a
longer baseline variability study.

\subsection{Apparent age spread and absolute age}

As was briefly discussed out in Section 1.3, the size of age spread inferred
from a given photometric spread depends on the median age of the star
forming region in question.  This should be borne in mind when
interpreting any unexplained spread in terms of an age spread.
For example, consider the $\sigma$ Ori young group.  If we accept an age of
5 Myrs and a distance of 350 pc, as we did in \citet{me2005}, then the 
remaining $R-i'$ spread of 0.062 mags represents a spread of approximately
4 Myrs (3.5 - 7 Myrs), based on the separation of NextGen isochrones
\citep{cb97,bcah2002} at $i' = 16.5$. 
If on the other hand we adopt the age found by
\citet{sww2004} of 2.5 Myrs and the distance they used for that estimate,
440 pc, we find the remaining $(R-i')$ spread corresponds to an age
spread of approximately 2 Myrs (2 - 4 Myrs).

In the case of the Cep OB3b association, \citet{pnjd2003} found the
ages of PMS objects to range from $< 1$ Myr to nearly 10 Myrs using
isochrones laid onto a $V / V-I$ CMD. 
As has been shown here, variability and binarity can only account for a small
fraction of this spread.

\section{Conclusions}
\label{sec:spreads_concs}

We have used 2 epoch, 2 colour photometry to investigate the influence
of photometric variability on the apparent age spreads in
CMDs.  We have found that the combination of binarity and variability
on timescales of $\sim$ years cannot account for the observed spread in
C-M space.  We argue that the remaining unexplained spread must either
reflect a genuine spread of ages, or longer timescale variability associated
with the changes in the accretion flow onto the PMS objects.

\bibliography{refs}
\bibliographystyle{mn2e}

\end{document}